# The fractal structure of the ventral scales in legless reptiles

**H A Abdel-Aal[1] M El Mansori[2]**

[1]*Department of Mechanical Engineering, Drexel University*, 3141 Chestnut St, Philadelphia, Pennsylvania, USA, 19144.
E-mail: hisham.abdelaal@Drexel.edu
[2]*Mechanics, Surfaces and Materials Processing Laboratory* (MSMP), Arts et Métiers ParisTech, 2 cours des Arts et Métiers, 13617, Aix-en-Provence, France

**Abstract.** Surface constructs in snakes reflect desirable design traits for technical surface engineering. Their micro-textural patterns, however, do not lend themselves easily to unified analysis due to species-specific variations in surface geometry and topology. Fractal description is useful in this context since it accentuates the correspondence between patterns especially when responding to tribological phenomena. In this work we examine the surface construction of 14 snake species, representing five families, and evaluate the fractal dimension for each of the skins (both the dorsal and ventral sides) using three different computational algorithms. Our results indicate first that all of the examined species share a common fractal dimension (with a very small variation between species in the order 4-5%). This finding implies that despite the different micro-geometry of texture among species, the skin as a unit responds in a similar manner to many interfacial influences.

**Introduction**
In the natural world, emphasis on micro-texture arises in response to the severe demands on efficiency, economy of effort, and on survival. In legless locomotion, not only the reptile has to economize its effort when moving, but also it has to utilize friction (dynamic or static) to generate tractions, balance motion, and to initiate propulsion [1-7]. These demands lead to a skin made essentially of the same chemical elements, yet with different local concentrations to meet local functional requirements (i.e., a skin with inherited compositional heterogeneity and added complexity). Frictional surfaces in snakes employ micro-texture as a means to compensate for this heterogeneity, and thereby manipulate system complexity, to optimize function. In this context pattern and distribution of topographical features work to facilitate the manipulation of system complexity.
General composition of surfaces within legless locomotors, entails textural elements of primitive geometry. The pattern, size, structural density, and spatial arrangement of these elements work holistically to facilitate multitasking. Deliberate spatial arrangement rather than the geometry of the textural elements appear to be of more prominence in engineering of these surfaces. Reduction of complexity through micro-texture takes place through two main mechanisms. The first is repetition, of micro-textural features of simple geometries and strategic spatial placement of features in a manner compatible with the multifunctional requirements of the surface. Adopting this elementary design philosophy is clearly desirable within the technological realm. However, there is a need to understand the sense of pattern formation, and topological distribution, within natural surfaces in light of the design and functional imperatives on the reptile. A point of entry to this endeavour is to study the unifying characteristics among the various textural patterns observed in snake surfaces. Micro-textural patterns however, do not lend themselves easily to the proposed unified analysis due to species-specific variations in surface geometry and topology.

---

[1]  To whom any correspondence should be addressed.

Due to the variations in geometry and other parameters, a unifying design concept may not be directly apparent. This hinders probing the surface to deduce a unified design principle. One remedy to this problem is to view the geometry, and topology, of a natural surface within a different frame of reference. This frame of reference should allow the abstract description of the surface (thus neutralizing the effect of variation in conventional geometry and associated effects). One powerful method useful in this context is fractal description.

Fractals efficiently describe the geometry of regular shapes that are not amenable to conventional description through Euclidean geometry. In particular, fractals describe how much space an irregular object occupies [8]. The structure of the fractal number is indicative of the nature of the object described. The higher the fractal dimension, the more space the curve occupies. That is, the fractal dimension (FD) indicates how densely a phenomenon occupies the space in which it is located. Furthermore, a fractal dimension is a unique number that is independent of the alteration of space through manipulation by stretching or compacting.

Two surfaces of the same fractal dimension do not necessarily manifest identical microstructure. Rather, they display a common attitude in their dependence on their roughness to occupy a three dimensional space. By focusing on the growth of form, and its evolution in space, fractal description of an object shifts the focus from devising the Euclidean geometry of an object to describing the evolution of its growth into its prescribed space. That is, fractal description of a surface deals with the evolution of the process of surface growth into the space rather than finding its boundaries with respect to an origin. As such, a fractal frame of reference projects differences within the micro-textural elements, comprising the surface, in a view that unmasks their common phenomenal traits. This accentuates the correspondence between behavioural patterns, of apparently dissimilar Euclidean form, especially when responding to tribological phenomena (e.g. contact forces or diffusion through surface topography). It follows that two surfaces sharing the same fractal dimension, in principle, should manifest common dynamic responses to interfacial events. To reveal the sense of pattern in micro-texture of reptilian skin, it is, therefore, essential to explore the fractal structure of the respective surface construction.

In this work, we attempt to find the common attributes of micro-textural patterns in snakeskin through examination of topology within a fractal frame of reference. To this effect, we examine the surface construction of 14 snake species representing five families of snakes to determine their fractal dimension (both the dorsal and ventral sides). We further, use the information extracted from the skins to point out the common behavioural aspects that should persist across species (especially with respect to locomotion and frictional response).

**2.0   Materials and methods**

*2.1   Species*

Allometry of size guided the choice of the species examined in this work. In choosing the species, we attempted to diversify our choice of species to reflect the wide variation of size and length in snakes. To this end, we examined the exuviae of the snakes shown in Table 1 (n=5). Table 2, meanwhile, provides a summary of the basic dimensions (mass and length) and ventral scale counts of the species.

Table-1 Summary classification and taxonomy of species examined in the current work

|  | Family: | Subfamily: | Genus: | Species: | Length cm | Mass (g) |
|---|---|---|---|---|---|---|
| Bittis Gabonica | Bitis | Viperinae | Bitis | B. gabonica | 150 | 1306 |
| Echis Carinatus | Viperdae | Viperinae | Echis | E. Carinatus | 72 | 142 |
| Agkistrodon Taylori | Viperidae | Crotalinae | Agkistrodon | A. bilineatus | 87 | 256 |
| Cerrophidion Godmani | Viperidae | Crotalinae | Cerrophidion | C. godmani | 84 | 227 |
| Agkistrodon Contortrix | Viperidae | Crotalinae | Agkistrodon | A. contortrix | 135 | 950 |
| Trimeresurus tokarensis | Viperdae | Crotalinae | Trimeresurus | T. Tokarensis | 125 | 755 |
| Agkistrodon piscivorus | Viperidae | Crotalinae | Agkistrodon | A. piscivorus | 154 | 1413 |
| Pseudechis Austarlis | Elapidae |  | Pseudechis | P. australis | 275 | 8144 |
| Gongylophis colubrinus | Boidae | Erycinae | Gongylophis | G. colubrinus | 91 |  |
| Epicrates Cenchria | Boidae | Boidae | Epicrates | E. Cenchria | 92 | 562 |
| Pituophis Melanoleucus |  | Colubrinae | Pituophis | P. melanoleucus | 154 | 1414 |
| Thamnophis Sirtalis | Colubridae | Natricinae | Thamnophis |  | 130 | 848 |
| L G. Californiae | Colubridae | Lampropeltis | L. Getula |  | 135 | 950 |
| Morelia Virdis | Pythonidae |  | Morelia | M.Virdis | 200 | 3113 |
| Python Regius | Pythonidae |  | Python | P. regius | 150 | 1306 |

*2.2  Characterization and computation of the fractal dimension*

All examined skin underwent three imaging procedures to extract detailed information about general dimensions, geometrical proportions, and essential metrological parameters. The procedure started by pre-identifying twelve equally spaced locations along the Anterior Posterior (AP) axis of each skin. SEM images, WLI-interferograms, and AFM scans were obtained for the dorsal and ventral scales at each location.

This work used three methods to compute the fractal dimension of the examined skins: Cube counting [9-11], Triangulation [9], and the Variance method [11, 12]. For each species, we extracted the fractal dimension from SEM images AFM scans in topography mode, and WLI-images taken at twelve predetermined areas on the Dorsal and Ventral sides of the skins. Five sets of images were taken randomly at different spots within each area for each skin side (dorsal and ventral). An image set consisted of an array of SEM imagery at various magnifications (250, 1000, 2500, 5000, 15000, and 50000 X), AFM scans of square areas (of sides 50, 25, and 10 μm respectively), and WLI-imagery that examined areas of dimensions 1250 μm by 500 μm. To each of the image sets we applied the three computational algorithms to extract the fractal dimension. In all, for each species, we analyzed 150 SEM images, 15 AFM scans, and 25 WLI-images for each predetermined spot using each of the three computational algorithms to estimate the respective fractal dimension.

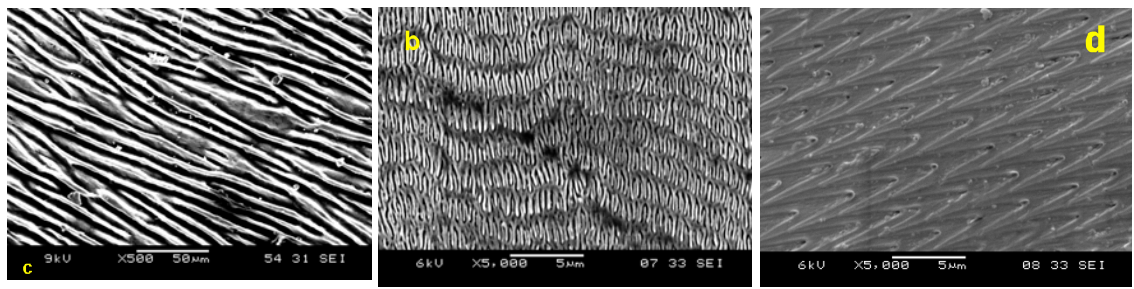

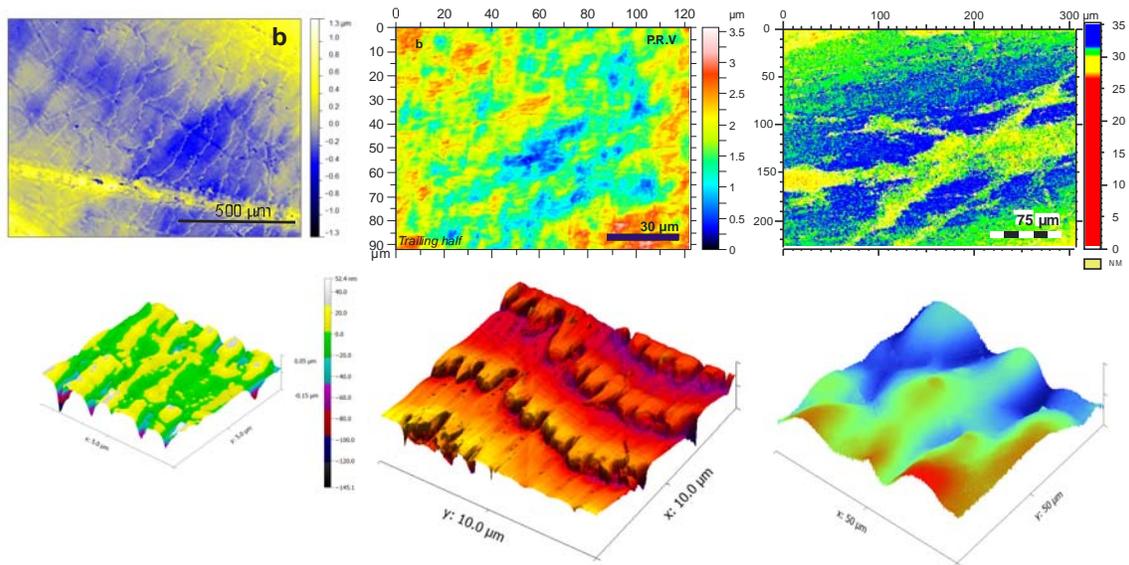

Figure 1 Representative imagery used to determine the fractal dimension of the skin. First raw (left to right) shows SEM imagery of the ventral scales of three snakes: C. Godmani, P. regius, and A. Piscivorus (X=5000). Second raw depicts white light interferograms for the same species. Third raw depicts AFM scans for the same species at different scan sizes.

## 3.0 Results and discussion
### 3.1 Results

Figure 2 (a and b) examines the variation in the fractal dimension (FD) at different zones within the body of two snake species P. regius and B. gabonica.

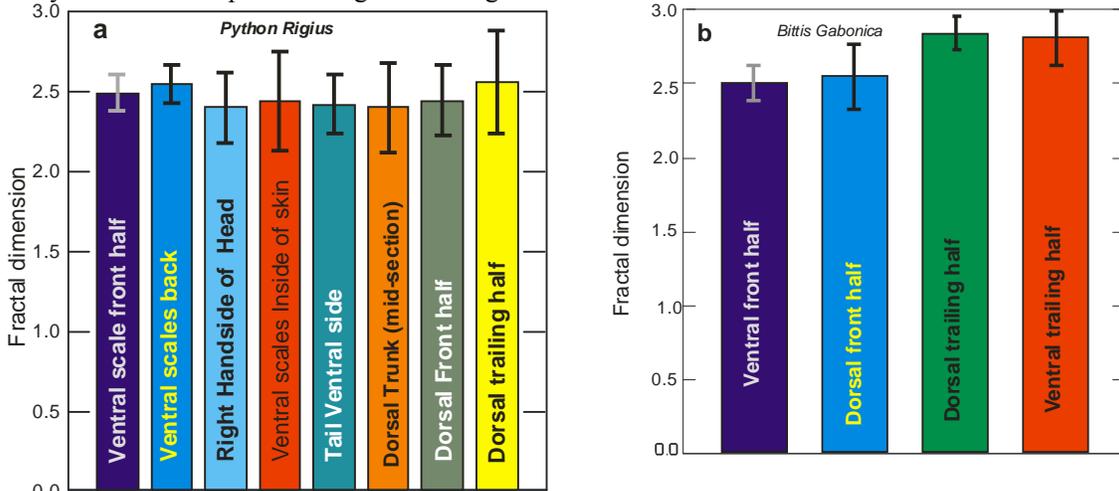

Figure 2 Comparison between local fractal numbers (dimensions) of the snakeskin examined in the current work. a- local fractal dimension of P. regius, b comparison of the fractal dimension of representative spots located on the ventral and dorsal sides of B. gabonica.

Figure 2-a, depicts the fractal dimension for four ventral spots and for dorsal spots on the Python species. The locations of these skin patches yield a fair representation of the fractal dimension within the leading and the trailing halves of the reptile. A remarkable observation of the plot is that, on average, the FD of all examined locations is almost equal (error percentage, however, differs by location). This implies that the surface of the species has a unique fractal value. Figure 2-b, which plots the fractal dimensions of four locations on the skin of B. Gabonica implies a different arrangement. Here the FD shows visible variation between examined body zones (and not sides i.e dorsal Vs ventral). The FD is equal for the ventral and dorsal skins on the front half of the body ($D_{FH} \approx 2.5$). A similar trend is observed for the trailing half of the body, however the FD differs from that of the leading body half ($D_{TH} \approx 2.765$). For comparison, $D_{FH}$ for B. Gabonica is almost equal to the average for P. regius (arithmetic mean of all values shown in figure 2-a).

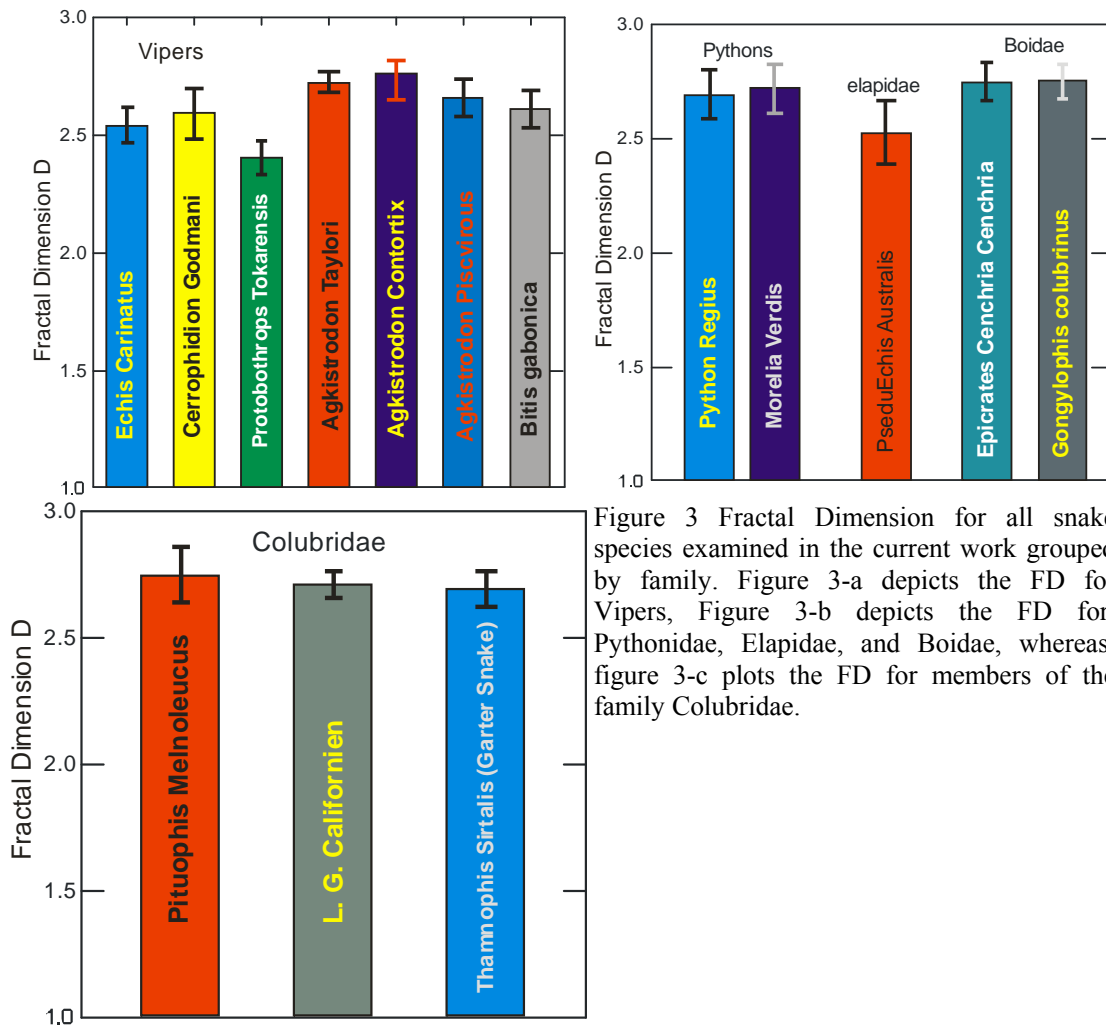

Figure 3 Fractal Dimension for all snake species examined in the current work grouped by family. Figure 3-a depicts the FD for Vipers, Figure 3-b depicts the FD for, Pythonidae, Elapidae, and Boidae, whereas, figure 3-c plots the FD for members of the family Colubridae.

Figure 3 (a-c) presents the average (arithmetic mean of the FD for all examined spots on a species) for all snakes examined in the current work. Each of the figures, when possible, presents the FD for a clad or family of snakes. Figure 3-a depicts the FD for Vipers, Figure 3-b depicts the FD for, Pythonidae, Elapidae, and Boidae, whereas, figure 3-c plots the FD for members of the family Colubridae. The data show that, in general, the FD is above 2.5 and close to three ($2.55 \leq D \leq 2.75$. Two species, P. Australis (Colubridae and P. Tokarensis (Viperdae), deviate from this value ($D \leq 2.5$).

The general trend of the data reveals a characteristic FD that distinguishes each genus examined. For example, within the genus Agkistrodon (A. Taylori, A. Piscvirous, and A Contortex) the FD is approximately 2.7, whereas for the genus Crotilanae, the FD falls around 2.55. This trend is more visible among Colubrides which manifest a FDas high as 2.75. This value is also shared by the family Boidae and the two python species examined.

*3.2   Discussion*

Discussion of the results will focus on implications for performance and surface construction rather than metrological aspects. The later discussion will be a subject of future work.

Fractals stand for a method to describe the geometry of regular shapes that are not amenable to conventional description through Euclidean geometry. In particular, fractals describe how much space an irregular object occupies [13]. A typical description of a surface is a so called Fractal number (dimension D). This number is a positive non-integer that varies between $1 \leq D \leq 3$. The structure of the fractal number is indicative of the nature of the object described. The most significant digit of the parameter D, stands for the topological dimension of the object, whereas the second part, the fractional part that varies from 0.0 to 0.999, is a so-called fractal increment. The higher the fractal dimension, the more space the curve occupies. That is, the fractal dimension indicates how densely a phenomenon occupies the space in which it is located.

A surface of which FD is close to 3 implies roughness that densely occupies its volume, and by so doing, the roughness manifests complex branching. The complexity resulting from the manner of filling the limited volume maximizes the pathways necessary for many functions (e.g. thermal regulation). Thus in essence, a high FD enhancing efficiency of particular skin functions. A high FD, close to three, may also imply a high ratio of surface area to volume. However the increase in area does not take place through extending a feature but through fractal branching of bends and folds of the skin. Physically, this indicates that the increased surface area is subdivided into quasi-independent hierarchical domains, which potentially can display swarm-like behavior. Hierarchical construction of textural features integrates the ability of the reptile to vary the contraction strength of ventral muscles to form an active control mechanism of friction tractions. This mechanism essentially can vary the propulsive forces necessary for motion and therefore controls the power requirements of the reptile.

An interesting finding of the analysis, figure 2-a, is the higher FD for the trailing body half of B. Gabonica than that of the leading half. Such an increase indicates that the ratio o f the volume occupied by the branching textural features is higher for the trailing half of the body which in turn implies functional adaptation.  B. Gabonica is a heavy ground dweller that inhabits regions with high rain fall.  Defense and feeding requirements mandate rigidity of the trailing half of the body. More roughness is required, whence the higher fractal branching of texture features.  On the other hand, the almost uniform FD for P. regius implies uniform density of roughness branching which helps function of the reptile within its arboreal habitat that may require the snake to cling to a tree trunk for example with the entire body stretched. Uniform fractal branching contributes even loading on ventral muscles (which contract to provide gripping forces) in resistance to gravitational tractions acting at the interface of the body and the clinging medium.

Furthermore, a fractal dimension is a unique number that is independent of the alteration of space through manipulation by stretching or compacting. Therefore, in essence, two surfaces that share a fractal dimension, share also several intrinsic behavioral characteristics. Mainly, the two surfaces share the manner by which they occupy their space (even if at first glance they appear fictitiously different) along with the consequences for accommodation of loads and transportation flux.

The skin being a visco-elastic material exhibits stress relaxation (i.e., an applied stress will vanish asymptotically to a zero state if the time duration of applying that stress is almost infinite (or sufficiently long)). Roughness of the surface influences relaxation behavior of the skin. It is possible to express the topology of the roughness in terms of the fractal dimension D. Within such formulation, the value of the FD being indicative of roughness density, may also indicate whether bulk deformation or roughness deformation would dominate the stress relaxation of the material and vice-versa. Any time stress is greater for the smaller FD, at the same temperature and same strain, bulk deformation will dominate. This is implied from the definition of the FD itself; lower fractal dimensions mean less roughness density; consequently, the bulk material dominates, while a higher FD indicates denser roughness and consequently asperity deformation dominates [14]. As such, for all snakes having close values of the FD imply also that they also share similar stress relaxation behavior and whether or not roughness deformation dominates such a process. The FD being very close to three implies the dominance of roughness deformation. This result is important as it points at the similarity of accommodating contact stresses applied to the skin (despite the apparent differences in ornamentation, size, and geometry of textural features).

Fractal surface geometry significantly impacts functional performance of materials. Friction of snakeskin is a function of contact stiffness. Bauem et. al., [15] reported that the coefficient of friction (COF) of the California king snake L. G. Californae depends on the roughness of the substratum as well as on the stiffness of the contact. In their experiments they reported that the COF of a rough glass ball on the skin in the cranial direction was some 30% lower than the COF of a smooth sphere. The variation was directional (for example, the lateral COF was some 12% higher for a rough sphere than for a smooth sphere sliding at identical conditions). Stiffness of the underlying skin layers also affects friction un-cushioned samples had considerably higher COF. Contact stiffness is a strong function of the surface roughness of the skin, so the density of roughness can explain that counter-intuitive behavior.

The FD of L. Californae is approximately 2.75 which implies dense roughness within the volume allowed for the skin. Rougher solid surfaces are more compliant and penetrable, their properties depend on surface morphological parameters. Denser roughness, implied by a high FD, causes a decrease of stiffness. Buzio *et. al*., [16] report that an increase of the FD by 10% for example causes a decrease in stiffness by one order of magnitude. The presence of denser roughness causes the reduction of the contact stress as well, which reduces the area of true contact and causes a lower friction force (thereby lower COF). The results of Pohrt and Popov [17] who studied the role of fractal roughness on contact stiffness support the proposed role of roughness in relation to the fractal structure of the studied skins.

Within classical contact models [18, 19] the apparent area of contact grows larger with added roughness compared to a smooth Hertzian case. The contact radius is proportional to contact stiffness in concentrated contacts. As such, within a classical formulation, one could expect the contact stiffness to grow. However, Pohrt and Popov point that this behavior is true beyond a critical load (which is a function of the Holder exponent and thereby the FD). For loads smaller than this transitional limit, added roughness leads to a decrease in contact stiffness (compared to a smooth surface), whence larger contact radius would lead to a lower contact area and thereby lower friction force.

A significant consequence of the common fractal dimension pertains to transdermal diffusion of particles. Pazkossy and Nykos [20] studied the decay of the diffusion of a current of particles diffusing from an initially homogenous medium to a completely absorbing fractal boundary. They found that such a process, which is important in studying of drug administration and absorbance, exhibits a time dependency of the form t-b as compared to the conventional time factor $t^{1/2}$. The exponent $\beta$ in this case is a direct function of the FD, and assumes the form *$\beta=0.5 (FD-1)$*. This corresponds to a generalized Cottrell equation and may be used to describe the frequency dispersion induced by roughness. Sharing a fractal dimension, therefore, indicates that all skins examined types have similar diffusion decay kinetics. This typically explains the clinically observed permeation compatibility between human and snake skins [21-23].

**Conclusions**

This work presented a comprehensive study of the fractal structure for snakeskin. Results indicated that the fractal dimension of snakes occupies a narrow band (around 2.75 ± 0.05). This implies that the skin, despite differences in shape and geometry of the micro-texture elements responds in a similar manner to interfacial phenomena (e.g., tangential and contact loading).

**References**
[1] Arzt, E., Gorb, S., Spolenak, R. (2003). From Micro to Nano Contacts in Biological Attachment Devices, *Proceedings of National Academy of Sciences (PNAS)*, 100 (19), 10603-10606.

[2] Hazel, J., Stone, M., Grace, M.S., Tsukruk, V. V. 1998 Tribological design of biomaterial surfaces for reptation motions. *Polymer Preprints*, 39, 1187-1188.
[3] Berthé, R. A., Westhoff, G., Bleckmann, H., Gorb, S. N., 2009. Surface structure and frictional properties of the skin of the Amazon tree boa Corallus hortulanus (Squamata, Boidae) *Journal of Comparative Physiology A: Neuroethology, Sensory, Neural, and Behavioral Physiology*, 195, 311–318.
[4] Abdel-Aal, H. A., El Mansori, M. 2013. Tribological analysis of the ventral scale structure in a Python regius in relation to laser textured surfaces, *Surface Topography and Metrological Properties*. 1 015001. doi:10.1088/2051-672X/1/1/015001
[5] Abdel-Aal, H. A. 2013. On Surface Structure and Friction Regulation in Reptilian Locomotion, *Journal of the Mechanical Behavior of Biomedical Materials* 22:115-135, DOI 10.1016/j.jmbbm.2012.09.014
[6] Gray, J., Lissemann, H. W., 1950. The kinetics of locomotion of the grass-snake. J. exp. Biol. 26: 354–367
[7] Pough, F. H., Groves, J. D. 1983. Specialization of the body form and food habits of snakes. *American Zoologist*, 23, 443–454.
[8] D.S. Ebert, F. K. Musgrave, D. Peachey, K. Peril, S. Worley. 1998. *Texturing and Modeling: A Procedural Approach*. Second Ed. Academic Press, San Diego.
[9] W. Zahn, A. Zösch: 1999 The dependence of fractal dimension on measuring conditions of scanning probe microscopy. *Fresenius J Analen Chem,* 365: 168-172
[10] W. Zahn, A. Zösch: Characterization of thin film surfaces by fractal geometry. *Fresenius J Anal Chem* (1997) 358: 119-121
[11] C. Douketis, Z. Wang, T. L. Haslett, M. Moskovits 1995 Fractal character of cold-deposited silver films determined by low-temperature scanning tunneling microscopy. *Physical Review B*, 51, 16,
[12] P. G. de Gennes, Physique des surfaces et des interfaces, C. R. *Acad. Sc. Paris* 295 (1982) 1061–1064.
[13] Osama M. Abuzeid and Taher A. Alabed 2009: Mathematical modeling of the thermal relaxation of nominally flat surfaces in contact using fractal geometry: Maxwell type medium, *Trib. International*, 42(2), 206-212.
[14] Baum M. J., Kovalev A. E., Michels J. and Gorb S. N. 2014 Anisotropic Friction of the Ventral Scales in the Snake Lampropeltis getula californiae *Tribology Letters* 54, 2, 139-150.
[15] Renato Buzio, Corrado B, Biscarini F, Buatier De Mongeot F, Valbusa U, 2003, The contact mechanics of fractal surfaces, *Nature Materials* 2, 233 - 236
[16] Pohrt R, Popov V, 2013 Contact Mechanics of Rough Spheres: Crossover from Fractal to Hertzian Behavior *Advances in Tribology*, 2013. doi:10.1155/2013/974178
[17] J. A. Greenwood and J. B. P. Williamson, Contact of nominally flat surfaces, *Proc of the Royal Society A,*. 295, 300–319, 1966
[18] J. A. Greenwood, J. H. Tripp, 1967 The elastic contact of rough spheres, ASME J. Applied Mechanics, 34, 1, 153–159.
[19] Tamás Pajkossy, Lajos Nyikos, 1989 Diffusion to fractal surfaces—II. Verification of theory, *Electrochimica Acta,* 34, 2, 171-179.doi./10.1016/0013-4686(89)87082-3.
[20] Bhatt, P.P., J.H. Rytting, E.M. Topp, 1991. Influence of azone and lauryl alcohol on the transport of acetaminophen and ibuprofen through shed snake skin., *Int. J. Pharm.*, 72: 219-226.
[21] Godin B , Touitou E 2007 Transdermal skin delivery: predictions for humans from in vivo, ex vivo and animal models *Adv Drug Deliv* Rev 59 (11):1152-61
[22] Harada, K., T. Murakami, E. Kawasaka, Y. Higashi, S. Yamanoyo and N. Yata, , 1993. In vitro permeability to salicylic acid of human, rodent and shed snake skin. *J. Pharm. Pharmacol.*, 45: 414-418.
[23] Haigh, J.M., E. Beyssac, L. Chanet and J.M. Aiache, 1998. In vitro permeation of progesterone from a gel through the shed skin of three different snake species. *Int. J. Pharm.*, 170: 151-156.